\documentclass[12pt]{article}

\usepackage{sbc-template}
\usepackage{graphicx,url}
\usepackage[utf8]{inputenc}
\usepackage[brazil]{babel}
\usepackage{verbatim}
\usepackage{listings}
\usepackage{xcolor}
\usepackage{caption}
\usepackage{subcaption}
\usepackage{fontawesome}

\usepackage{float}

\title{Towards 6G Network Slicing}

\author{
    Rodrigo Moreira\inst{1,2},
    Flávio de Oliveira Silva\inst{1}
}

\address{
    Faculty of Computing (FACOM) \\ Federal University of Uberlândia (UFU), Uberlândia, MG, Brazil
    \nextinstitute Institute of Exact and Technological Sciences (IEP) \\Federal University of Viçosa (UFV), Rio Paranaíba, MG, Brazil
    \email{\{rodrigo.moreira, flavio\}@ufu.br, rodrigo@ufv.br}
}

\begin{document} 

\maketitle

\begin{abstract}

Networks should connect communicating peers, supporting vertical services requirements. The network evolution towards 6G requires native network slicing techniques. Some literature approaches claim network slice realization, but they do not convincingly address the deployment across multiple Autonomous Systems. This work investigates the current 6G network slicing landscape, presents some gaps, and introduces the concept of the recursive network slicing between multiple Autonomous Systems, supported by the NASOR approach. This innovative concept supports implementing new network services required by the 6G vision. This work also sheds light on the 6G requirements for network slicing.

\end{abstract}
     
%\begin{resumo} 
%Edge Computing foi proposto para resolver problemas como gerenciamento de largura de banda, dispersão geográfica, permitindo o processamento de dados próximo aos usuários finais. O objetivo de Network Functions Virtualization (NFV) é simplificar o fluxo de trabalho para criação de serviços computacionais. Este trabalho avança o estado da arte ao propor e avaliar uma arquitetura de experimentação compatível com a plataforma de gerenciamento e orquestração ETSI para Virtualized Everything Functions (VxFs) em Edge Computing. Essa plataforma é capaz de implantar e monitorar continuamente VxFs para os mais diferentes experimentos, independentemente do hardware e a plataforma virtualização.
%\end{resumo}

\section{Introduction}\label{sec:introducao}

The purpose of networks is to connect communicating peers and provide their communication requirements. Networks' evolution requires their capacity to support different requirements of the top a common substrate. 

5G highlighted this by defining three generic services called Enhanced Mobile Broadband (eMBB), Massive Machine-type Communications (mMTCs), and Ultra-Reliable Low-Latency Communications (URLLCs) \cite{popovski_5g_2018} to support not only user-centric communication but also different business verticals. The trend of network specialization will continue and will require improvements in 6G \cite{dogra_survey_2020}.

A Network Slice \cite{ngmn2016description} is a logical network with specific characteristics, policies, and configurations. The network slice represents an on-demand, end-to-end logical network that runs on common hardware \cite{7926921} that supports user services and applications. The concept of network slicing is crucial to realize 6G vision \cite{chowdhury_6g_2020}.  

%//challenges

%//work objective
This work investigates the current 6G network slicing landscape, presenting some gaps that need to be fulfilled by network sling techniques to support the 6G vision. We also propose the concept of a recursive multi-domain network slice supported by the \textbf{N}etwork \textbf{A}nd \textbf{S}lice \textbf{OR}crhestrator (NASOR) approach for network slicing between multiple Autonomous Systems (AS).

The remaining of this work is organized as follows: Section~\ref{sec:related} presents an overview of the network slicing landscape regarding 6G. Section~\ref{sec:nasor} describes our proposal of a recursive multi-domain network slicing approach that couples to the 3GPP management system and represents an innovative concept of network slicing to support 6G requirements. Finally, in Section~\ref{sec:concluding_remarks} we draw some conclusions and present a future research agenda.

\section{The 6G Network Slicing Landscape}\label{sec:related}

\cite{9214395} proposes a vision for a 6G architecture based on blockchain with support for distributed and virtualized computing and communication for a decentralized multi-operator 6G environment. The proposal covers the actors and transactions related to a dynamic infrastructure and an electromagnetic spectrum market. This vision does not detail the technological components that enable network slicing. 6G architecture envisions the ability to deploy network slices driven by application needs and adaptable to specific markets~\cite{taleb2020white}.

%\cite{9214395} propôs uma nova visão para uma arquitetura 6G baseada em \textit{blockchain} com suporte para comunicação e computação distribuída e virtualizada para um ambiente 6G descentralizado multioperador. A proposta abrange os atores e transações relacionados com uma infraestrutura dinâmica e um mercado de espectro eletromagnético.

%Diferente da visão que este presente artigo propõe, a arquitetura de \cite{9214395} não possui componentes tecnológicos habilitadores de fatiamento de rede. Vislumbra-se para arquiteutra 6G a capacidade de implantar fatias de rede orientado pelas necessidades das aplicações e adapável para mercados específicos~\cite{taleb2020white}.

\cite{9003619} presents an architectural vision considering the advances of softwarization, agile control, and deterministic services over the 6G architecture. In addition, an architecture with modular components is envisaged to improve the customization and performance of network slicing. 

%\cite{9003619} apresentam uma visão arquitetural considerando os avanços da softwarization, controle ágil e serviços deterministicos sobre a arquiteura 6G. Além disso, prospecta-se uma arquitetura com componentes modulares para aprimorar a customização e realização de fatiamento de rede. 

%\cite{9003619} apresentam uma visão arquitetural considerando os avanços da softwarization, controle ágil e serviços deterministicos sobre a arquiteura 6G. Dentre as tecnologias habilitadores prospectadas para a nova aquitetura, espera-se a utilização ampla de millimeter wave, Multiple-input and Multiple-Output (MIMO) enhancements e cell-edge para amplificar a área de cobertura. Além disso, prospecta-se uma arquitetura com componentes modulares para aprimorar a customização e realização de fatiamento de rede. 

In contrast to ~\cite{9003619}, we argue that the realization of deterministic services, customization of network parameters at the transport level is achievable through what we call \emph{recursive slicing}, presented in Section \ref{sec:nasor}. 

%Diferente de~\cite{9003619}, vislumbramos que a realização de serviços deterministicos, customização de parâmetros de rede no nível do transporte é realizável por meio do fatiamento recursivo.

A proposed paradigm shift named \emph{Mandate-driven Networking} is presented by \cite{9083766}. To realize this paradigm, a hyperstrator, combines several network segments and administrative domains to carry out End-to-End (E2E) communication between humans or devices. In this proposal, it is considered that specific domains have their own orchestrators and that the requirements for the deployment of the network slice are expressed through the Expressive Application Program Interface (xAPI).

%Uma proposta de paradigm shifit nomeada Mandate-driven Networking é apresentada por \cite{9083766}. Essa proposta de fatiamento de rede considera combinar, por meio de um hiperstrator, vários segmentos de rede e domínios administrativos para realizar a comunicação E2E entre humanos ou dispostivos. Nessa proposta, considera-se que domínios específicos tenham seus próprios orquestradores e que os requisitos para a implantação da fatia de rede sejam expressos por meio da Expressive Application Program Interface (xAPI).

Our view of network slicing goes beyond the one presented by~\cite{9083766}. We envision network slicing as a fundamental technological enabler for the 6G architecture, extending its concept to recursive network slicing across multiple domains. By enabling recursive network slicing, numerous user or market-specific applications are achievable, such as ManyNets~\cite{taleb2020white}. 

%Nossa visão de fatiamento de rede vai além da apresentada por~\cite{9083766}. Vislumbramos o fatiamento de rede como um habilitador tecnológico fundamental para a arquitetura \textit{6G}, estendendo seu conceito para o fatiamento recursivo de rede recursivo entre múltiplso domínios. Ao habilitar o fatiametno de rede recursivo, inúmeras aplicações dos usuários ou específicas de mercado são realizáveis, como as \textit{ManyNets}~\cite{taleb2020white}.

In \cite{9200631} an architecture for the 6G era is discussed from the perspective of four aspects: platform, functions, orchestration, and specialization. Concerning specialization, 6G network architecture envisions offering specialized and performance attribute-oriented networks. This architectural block will contain mechanisms that perform extreme network slicing. Orchestration refers to components for offering open services, monetizing domain resources, and cognitive management and automation mechanisms.

%\cite{9200631} discute-se uma arquitetura para a era 6G sob a ótica de quatro aspectos: plataforma, funções, orquestração e especialização. Plataforma refere-se a característica \textit{het-cloud} que levará a arquitetura a operar de forma aberta, escalável, orientada a fluxos de dados, com aceleração de hardware e execução independente do ambiente. Funções refere-se a possibilidade da arquitetura oferecer convergência entre o acesso e o núcleo por meio de uma arquitetura de informação e Artificial Intelligence. 

%Especialização da arquitetura de redes de sexta geração prevê oferecer redes especializadas e orientadas a atributos de desempenho. Esse bloco arquitetural conterá mecanismos que realizam fatiamento de rede extremo. Orquestração refere-se aos componentes para oferta de serviço abertos, monetização de recursos do domínio e mecanismos cognitivos de gerência e automação.

The architectural vision of the 6G network proposed by~\cite{9200631}, specifically the realization of network slicing, differs from our vision. We envision a 6G network architecture using a distributed-hierarchical orchestrator to perform recursive slicing. However, the recursive network slicing proposed here can be improved through the concept of deep slicing, allowing a fine adjustment of the various parameters that describe a slice. 

%A visão arquitetural de rede 6G proposta por~\cite{9200631}, especificamente a realização de fatiamento de rede, difere da nossa visão. Vislumbramos para uma arquitetura de rede 6G um orquestrador distribuído-hierárquico para realização de fatiamento recursivo. Entretanto, o fatiamento de rede recursivo aqui proposto pode ser aprimorado por meio conceito de deep slicing, permitindoum ajuste fino dos diversos parâmetros envolvidos em um slice.

\cite{9083831} presents a view of user requirements taking into account the evolution of 5G. In addition, the author points out his perception of the essential technological resources and enablers for the realization of 6G 
%\cite{9083831} apresenta uma visão dos requisitos de usuário levando em conta a evolução do 5G. Além disso, o autor aponta sua percepção dos recursos e habilitadores tecnológicos essenciais para realização do 6G.  

In~\cite{9083831} it is considered that 6G network will provide dynamic network slicing. The 5G network has broadly addressed offering static and specialized connectivity as URLLC or eMBB. Our vision of network slicing expands the concept from dynamic to recursion. Thus, the owner of a network slice can dynamically resize it and offer subslices for specific applications and markets. 

%Em~\cite{9083831} considera-se que rede 6G proverá dynamic network slicing. A rede 5G amplamente abordou a oferta de conectivdiade estática e especializada como Ultra-Reliable Low-Latency Communication (URLLC) ou enhanced Mobile Broadband (eMBB). Nossa visão de fatiamento de rede expande o conceito de dinâmico para a recursividade. Assim, o dono de uma fatia de rede pode dinamicamente redimensioná-la e oferecer as subfatias para apliações e mercados específicos.

To summarize this brief landscape of network slicing for a 6G network, we propose Table~\ref{tab:short-landscape-table}. It presents characteristics such as \textbf{Proposal Type}, which refers to the type of contribution under analysis. Architecture Proposal contributions present components and technology enablers of a 6G network architecture. The \textbf{Recursive Network Slicing} feature denotes which work considers this concept. In this column, we distinguish our view of network slicing from other related work. 

%Para resumir o short landscape do fatiamento de rede para rede 6G, propomos a Tabela~\ref{tab:short-landscape-table}. Nela são apresentadas características como \textbf{Proposal Type}, que se refere ao tipo da contribuição em análise. Contribuiçõs que versam sobre Proposal Architecture apresentam componentes e habilitadores tecnológicos de uma arquitetura para rede 6G. A característica \textbf{Recursive Netowork Slicing} denota quais trabalhos oferecem esse tipo de feature, nessa coluna distinguimos nossa visão de fatiamento de rede dos demais trabalhos. 

Other columns such as \textbf{SDN}, \textbf{NFV}, \textbf{Blockchain} and \textbf{Artificial Intelligence} are the technological enablers that the views or architectures consider essential to perform a 6G network slice. The column \textbf{Multi-domain Orchestration} characterizes if the solution envisions or considers the architectural block components that enable network slicing between multiple domains. 

%Demais colunas como \textbf{SDN}, \textbf{NFV}, \textbf{Blockchain} and \textbf{Artificial Intelligence} são os habilitadores tecnológicos que as visões ou arquiteturas consideram essenciais para realização de fatiamento de rede 6G. A coluna \textbf{User Requirements} refere-se ao tipo de requisito de usuários que o fatiamento de rede é capaz de habilitar. A coluna \textbf{Multi-domain Orchestration} caracteriza se a solução vislumbra ou considera no bloco arquitetural componentes que habilitem o fatiamento de rede entre múltiplos domínios.

\begin{table}[h]
\centering
\caption{Network Slicing Proposals Summary.}
\label{tab:short-landscape-table}
\resizebox{\textwidth}{!}{%
\begin{tabular}{cccccccc}
\hline\hline
\textbf{Proposal} & \textbf{\begin{tabular}[c]{@{}c@{}}Proposal \\ Type\end{tabular}} & \textbf{\begin{tabular}[c]{@{}c@{}}Recursive \\ Network Slicing\end{tabular}} & \textbf{SDN} & \textbf{NFV} & \textbf{Blockchain} & \textbf{\begin{tabular}[c]{@{}c@{}}Artificial\\ Intelligence\end{tabular}} & \textbf{\begin{tabular}[c]{@{}c@{}}Multi-domain \\ Orchestration\end{tabular}} \\ \hline\hline
\cite{9214395}           & \begin{tabular}[c]{@{}c@{}}Architectural \\ Proposal\end{tabular} & \faCircleO                                                                    & \faCircle    & \faCircleO   & \faCircle           & \faCircle       & \faCircle                                                                      \\
%\hline
\cite{9003619}           & \begin{tabular}[c]{@{}c@{}}Architectural\\ Vision\end{tabular}    & \faCircleO                                                                    & \faCircle    & \faCircle    & \faCircleO          & \faCircle                                                                  & \faCircleO                                                                     \\
%\hline
\cite{9083766}           & Vision                                                            & \faCircleO                                                                    & \faCircleO   & \faCircle    & \faCircleO          & \faCircle                                                                  &  \faCircle                                                                      \\
%\hline
\cite{9200631}           & Vision                                                            & \faCircleO                                                                    & \faCircle    & \faCircle    & \faCircleO          & \faCircle                                                                 & \faCircle                                                                      \\
%\hline
\cite{9083831}           & Vision                                                            & \faAdjust                                                                      & \faCircleO   & \faCircle    & \faCircle           & \faCircle                                                                           & \faCircleO                                                                     \\
%\hline
\textit{Ours}              & \begin{tabular}[c]{@{}c@{}}Architectural \\ Proposal\end{tabular} & \faCircle                                                                     & \faCircle    & \faCircle    & \faCircleO          & \faCircleO                               & \faCircle                                                                      \\ \hline \hline
\end{tabular}%
}
\end{table}

In Table \ref{tab:short-landscape-table}, we point out how some state-of-the-art approaches adopt these several concepts. The symbol (\faCircle) represents the complete adoption. The (\faAdjust) suggests that the paper partially adopts the concept, the (\faCircleO) means the concept is not mentioned in work.

%In this paper, we point out how some state-of-the-art approaches fulfill the table requirements. The symbol (\faCircle) represents the complete feature realization. The (\faAdjust) suggests that the paper partially fulfills the requirements, the (\faCircleO) means the not fulfill requirement realization by the paper.

%\section{An Initial Approach for 6G Network Slicing}\label{sec:nasor}
%\section{Vision for 6G Network Slicing}\label{sec:nasor}
\section{Recursive Multi-Domain Network Slicing}\label{sec:nasor}

Network slicing has received significant efforts from the scientific community, specifically in the mobile networks context~\cite{7926923}. Recently, we proposed the NASOR approach for network slicing between multiple ASs~\cite{nasor}. 

NASOR is an architectural framework that employs political and technological separation of domains through hierarchical-distributed orchestrators. These Orchestrators interact with each other asynchronously through a key-value repository. Also, they exchange domain capacity information and network slice deployment parameters. Through segment routing technology, NASOR identifies network slices along the data plane constructed by Internet routing algorithms. 

\begin{figure}[!ht]
	\begin{center}
		\includegraphics[width=0.99\textwidth]{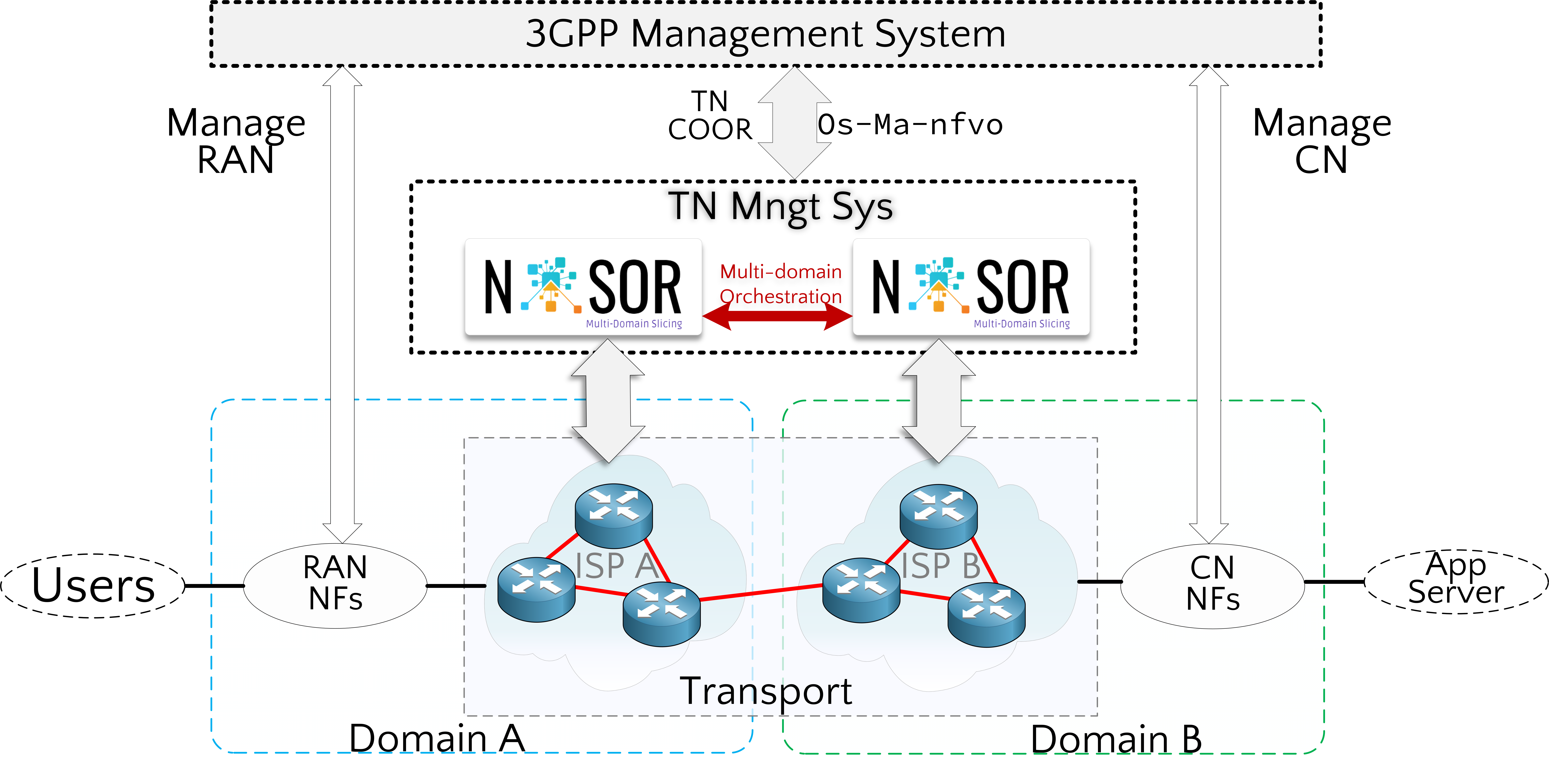}
	\end{center}
	\caption{Recursive Network Slicing Proposal at the Transport Level.}
	\label{fig:transport_network_slicing}
\end{figure}

There are many approaches in the state of the art which tried to realize network slicing. However, they predominantly focused on network slicing realization at mobile network access. As illustrated in Figure.~\ref{fig:transport_network_slicing}, we exploit network slicing to transport network level. A 3rd Generation Partnership Project (3GPP) leaves the transport level open to embed new technologies and management proposals and methods for network slicing realization at the transport level~\cite{3gpp}. Hence, NASOR plays Network Slice Management Function (NSMF) roles to handle network slicing life-cycle on the transport level.

In this paper, we glimpse the 3GPP Management System interacting with the NASOR through the \emph {Os-Ma-nfvo} interface to specify the network slice parameters. As the NASOR is hierarchical-distributed, the network slice deployment can span across multiple ASs. The NASOR builds a network slicing through the Network and Orchestration (NANO) component. NANO is the private orchestrator for a new network slicing. NANO enables the users or slice-owner to configure theirs network slicing according to the desired parameters, such as latency, throughput, or resize the network slice itself, creating new sub-slices. Rescaling a parent network slice by generating subslices of child networks is a recursive network slice. 

\begin{figure}[!ht]
	\begin{center}
		\includegraphics[width=0.95\textwidth]{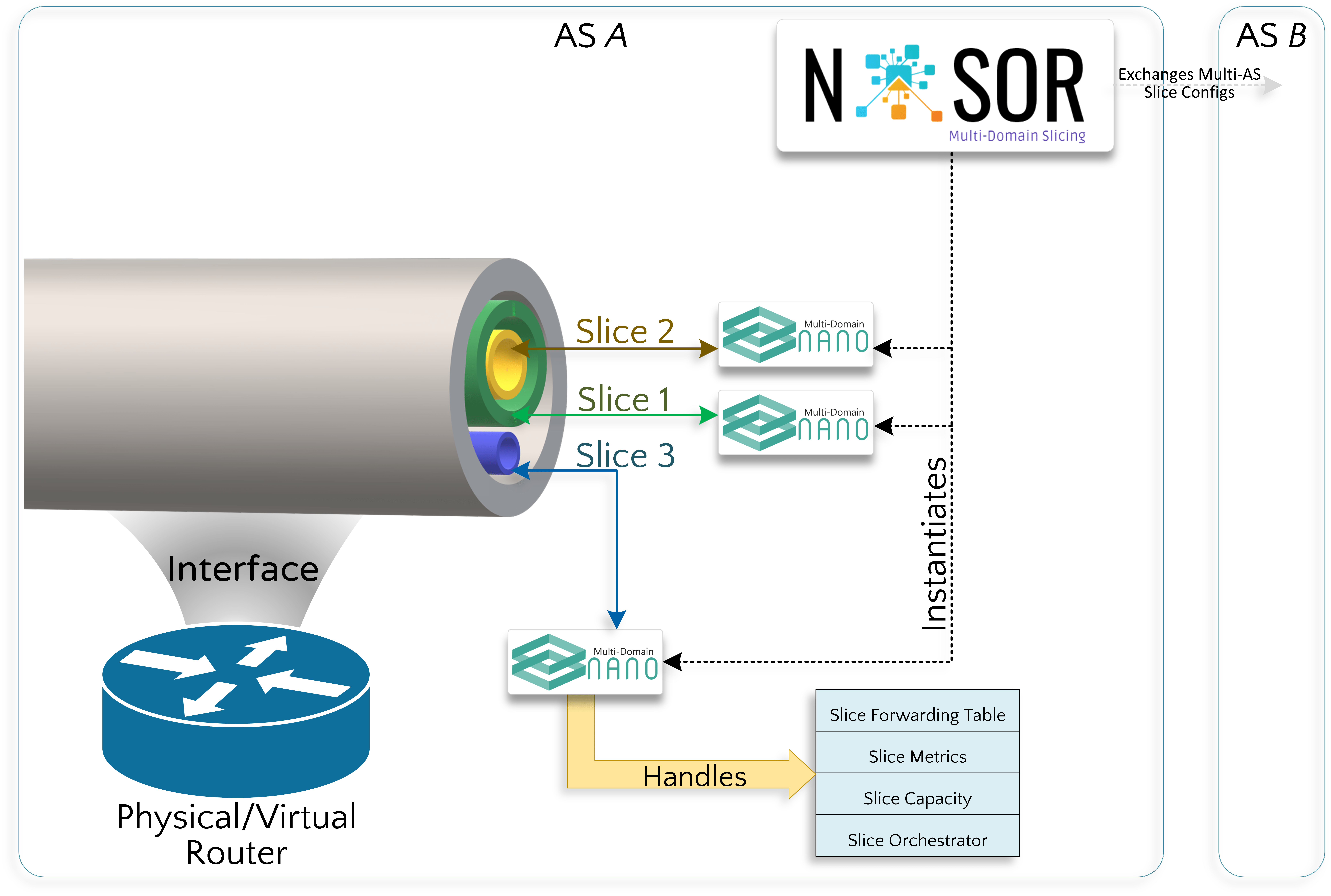}
	\end{center}
	\caption{Recursive Multi-domain Network Slicing.}
	\label{fig:recursive_network_slicing}
	%\legend{Fonte: Adaptado de}
\end{figure}

Figure.~\ref{fig:recursive_network_slicing} depicts how NASOR employs recursive network slicing for 6G networks. Upon receiving a network slice deployment request, NASOR instantiates a new NANO that will handle network slicing configuration and orchestration. NANO configures parameters across Internet routers to pave a data plane for the network slice. NANO sets up the Slice Forwarding Table, which contains the network slices identifiers. NANO handles Slice Metrics, defines Slice Capacity. According to Fig.~\ref{fig:recursive_network_slicing}, NANO enables slice-owner to create new slices, Slice 1 contains a Slice 2, which was recursively deployed. Eventually, Slice 2 may resize itself, creating a network subslice up to their parent network slicing resource limit.

The 6G architecture should enable more sophisticated use cases and meet stringent requirements such as latency and reliability. Thus, among the technological and conceptual enablers for performing network slicing, a recursive multi-domain network slicing mechanism emerged as a fundamental enabler. In addition, intelligent network slicing should shift paradigms with customized network slices that are adaptable to user demands. 

\section{Concluding Remarks and Future Work}\label{sec:concluding_remarks}

Network slicing is critical to serving users in a personalized way. The \emph{one-size-fits-all} concept makes innovative applications and business verticals over the network impossible. So, in this work, we present a brief landscape of network slicing regarding the 6G vision. 

We also present NASOR, a framework for deploying network slices across multiple domains, and supports the concept of recursive network slicing. NASOR is a subsystem that interacts with the 3GPP Management System to perform recursive slicing at the transport level. Furthermore, we pointed out how the recursive slicing, enabled by NASOR, is an innovative concept to support the implementation of new network services required by the 6G vision.

%O fatiamento de rede é fundamental para atender aos usuários de forma personalizada. O conceito ``one-fits-all'' impossibilita inovative applications e verticais de negócios sobre a rede. Assim, nesse artigo apresentamos um short landscape do fataiemtno de rede na esteira do 6G e propusemos o NASOR. O NASOR é um framework de implantação de fatias de rede entre múltiplos domínios. Posicionamos o NASOR como um componente que atua como um subsistema do 3GPP Management System para realizar o fatiamento recursivo no nível do transporte. Alem disso, pontuamos como o fatiamento recursivo, habilitado pelo NASOR, será capaz dar um paradigm shift na concepção e implantação de novos serviços de rede.

In our research and development agenda, we intend to evaluate the recursive network slicing quantitatively, considering scalability, security, and isolation challenges.
Furthermore, it seems critical to assess how artificial intelligence techniques can expand the possibilities regarding managing and intelligently orchestrating self-sustained recursive network slices considering, among others, smart slicing provisioning and performance assurance awareness.

%Como trabaho futuro, vislumbramos avaliar quantitativamente o fatiamento de rede recursivo. Além disso, parece-nos fundamental avaliar como as técnias de inteligência artificial podem expandir as possibilidades de gerenciamento e orquestração de fatias de rede para habilitar consumo eficiênte de energia.

\section*{Acknowledgement}\label{sec:agradecimentos}

This study was financed in part by the Coordenação de Aperfeiçoamento de Pessoal de Nível Superior - Brasil (CAPES) - Finance Code 001.

\bibliographystyle{sbc}
\bibliography{sbc-template}

\end{document}